\documentclass[final,11p,times]{article}
\usepackage{verbatim}
\usepackage[a4paper,margin=2.54cm]{geometry}
\usepackage{float}
\usepackage{authblk}
\usepackage{amssymb}
\usepackage[english]{babel}
\usepackage[version=4]{mhchem}

\usepackage{amsmath}
\usepackage[superscript,biblabel]{cite}
\usepackage{graphicx}
\usepackage[colorlinks=true, allcolors=blue]{hyperref}
\usepackage{subfigure}
\usepackage{subcaption}
\usepackage{caption}
\usepackage{siunitx}
\DeclareSIUnit{\ions}{ion}
\usepackage{miller}
\usepackage[nolist]{acronym}
\usepackage{setspace}

\renewcommand{\abstractname}{\normalsize Abstract}  
\makeatletter
\renewenvironment{abstract}{
    \small          
    \begin{center}  
    \bfseries        
    \large \textnormal{\abstractname} 
    \end{center}
    \quote
}
{\endquote}
\makeatother

\newcommand{\bgao}{$\beta$-\ce{Ga2O3}}
\newcommand{\ggao}{$\gamma$-\ce{Ga2O3}}
\DeclareSIUnit\bar{bar}
 \DeclareSIUnit\angstrom{\text {Å}}
\begin{document}
\setstretch{2}
\title{\textbf{Effect of Near-surface Thermal Spikes on Radiation Hardness of Gallium Oxide}}

\author[1]{Tomás Fernández Bouvier\thanks{Corresponding author: tomasfbouvier@gmail.com}}
\author[2]{Umutcan Bektas}
\author[3]{Alexander Azarov}
\author[1]{Ru He}
\author[2]{Nico Klingner}
\author[2]{René Hübner}
\author[4]{Paul Chekhonin}
\author[1]{Aleksi Leino}
\author[1]{Kai Nordlund}
\author[3]{Javier García Fernández}
\author[3]{Andrej Kuznetsov}
\author[2]{Gregor Hlawacek}
\author[1]{Flyura Djurabekova}
\affil[1]{Accelerator Laboratory, University of Helsinki. Pietari Kalmin katu 2, Helsinki, 00560, Finland}
\affil[2]{Institute of Ion Beam Physics and Materials Research, Helmholtz-Zentrum Dresden-Rossendorf, Bautzner Landstraße 400, Dresden, 01328,Germany}
\affil[3]{University of Oslo, Centre for Materials Science and Nanotechnology, PO Box 1048 Blindern, Oslo, 0316, Norway}
\affil[4]{Institute of Resource Ecology, Helmholtz-Zentrum Dresden-Rossendorf, Bautzner Landstraße 400, Dresden, 01328,Germany}

\begin{acronym}
  \acro{dft}[DFT]{Density Functional Theory}
  \acro{dpa}[dpa]{displacements per atom}
  \acro{ebsd}[EBSD]{Electron Backscatter Diffraction}
  \acro{fft}[FFT]{Fast Fourier Transform}
  \acro{fib}[FIB]{Focused Ion Beam}
  \acro{hzdr}[HZDR]{Helmholtz-Zentrum Dresden-Rossendorf}
  \acro{hrtem}[HRTEM]{High-Resolution \acl*{tem}}
  \acro{tem}[TEM]{Transmission Electron Microscopy}
  \acro{saed}[SAED]{Selected Area Electron Diffraction}
  \acro{stem}[STEM]{Scanning \acl*{tem}}
  \acro{sem}[SEM]{Scanning Electron Microscope}
  \acro{bfstem}[BF-STEM]{Bright-Field \acl*{stem}}
  \acro{lmais}[LMAIS]{Liquid Metal Alloy Ion Source}
  \acro{rbs}[RBS]{Rutherford Backscattering Spectrometry}
  \acro{rbsc}[RBS/c]{\acl*{rbs} in channeling mode}
  \acro{haadf}[HAADF]{High-Angle Annular Dark-Field}
  \acro{haadfhrtem}[HAADF-HRTEM]{High-Resolution \acl*{tem}}
  \acro{haadfstem}[HAADF-STEM]{High-Angle Annular Dark-Field \acl*{stem}}
  \acro{ptm}[PTM]{Polyhedral Template Matching}
  \acro{bca}[BCA]{Binary Collision Approximation}
  \acro{fcc}[FCC]{Face Centered Cubic}
  \acro{bf}{Bright-Field}
\end{acronym}

\maketitle
\begin{abstract}
Gallium oxide (\ce{Ga2O3}) stands out as an extraordinary high radiation-tolerant semiconductor, because its lattice displacements induce polymorph transitions, holding material crystalline instead of leading to amorphization. Meanwhile, under extremely severe irradiation conditions many crystals become amorphous, often starting from surfaces where the translation symmetry breaks. Here, we show that the surface amorphization may prevail over the crystallisation in \ce{Ga2O3}, however only if the mass and energy of irradiated ions produce sufficiently dense heat spikes in the immediate vicinity of the free surface. Applying machine-learned molecular dynamics simulations together with experimental broad-beam and focused ion-beam irradiations, we conclude that the presence of the free surface enables asymmetric displacements of Ga and O atoms, leading to a local non-stoichiometry. Consequently, when the affected cascade volume is sufficiently large, this compositional imbalance suppresses recrystallization and promotes amorphization. As such, our results are ready to use for tailoring irradiation conditions to either prevent or induce surface amorphization, depending on the requirements of the intended applications in \ce{Ga2O3} or other compound semiconductors.
\end{abstract}
\section*{\underline{\makebox[\textwidth][l]{Introduction}}}
The radiation hardness in solids, particularly in semiconductors, is critical to enabling technologies to operate in harsh radiation environments \cite{Prinzie2021}. Importantly, gallium oxide (\ce{Ga2O3}) – a prominent ultra-wide bandgap semiconductor – has recently been discovered to exhibit extraordinarily high radiation tolerance in the keV-energy ion irradiation regime \cite{azarov_universal_2023}. This phenomenon has been explained in terms of disorder-induced ordering in the thermodynamically stable monoclinic phase of \ce{Ga2O3} (known as $\beta$-\ce{Ga2O3}), so that instead of amorphization, a transition to another polymorph is more energetically favorable \cite{Azarov2022}. This newly formed polymorph has now been unambiguously identified as the cubic defective spinel (known as $\gamma$-\ce{Ga2O3}) \cite{PhysRevLett.134.126101} and has been found to withstand fluences far beyond those tolerated by other semiconductors, essentially without loss of crystallinity \cite{azarov_universal_2023}. This behavior was attributed to the exceptional resilience of oxygen sublattice under irradiation conditions \cite{PhysRevMaterials.8.084601}. 
In view of this, it is natural that observations of \ce{Ga2O3} amorphization are rare. Nevertheless, Azarov et al. \cite{azarov_universal_2023} reported that the amorphization of \bgao~ could occur because of deviations from stoichiometry, e.g. introducing high contents of matrix Ga or O elements or chemically active impurities, both typically localized in near-surface regions. Lorenz et al. \cite{Lorenz2014} observed surface-initiated amorphization under high fluences, while Azarov et al. \cite{Azarov2025} reported surface amorphization of the orthorhombic polymorph surface ($\alpha$-\ce{Ga2O3}) prior to formation of the $\gamma$-phase, attributed to the accumulation of excess tensile stress. These findings collectively highlight the critical role of the surface in amorphization processes in this class of materials otherwise highly radiation tolerant. 

Generally speaking, amorphization is governed by the instant increase in defect density that is maintained in the region of interest, e.g. near the surface. In conventional broad ion beam (BIB) experiments, when ion impacts are widely separated in time and space, there is sufficient time for lattice recovery 
between two consecutive displacement cascades that may occur in the same location.
In contrast, much higher ion fluxes during focused ion beam (FIB) implantation increase the probability of displacement cascade overlap, increasing the defect density, subsequently impeding the lattice recovery, and accelerating the amorphization of the structure. 
Furthermore, it has been shown that the density of displacement cascades induced by ions of varying masses may also constitute a decisive parameter for the onset of amorphization
\cite{Titov2025, Karabeshkin2022, Struchkov2024,Sarwar2024}. For example, Klevtsov et al \cite{Klevtsov2024} suggested that heavy monoatomic or molecular ions can induce thermal spikes within the \ce{Ga2O3} lattice, triggering stoichiometric imbalance and facilitating amorphization instead of the formation of the $\gamma$-phase.

To determine which of the processes listed above is responsible for amorphization in radiation-resistant material,
we performed a systematic study that combines multiscale computational modeling with multirange ion irradiation experiments. 
Our experiments included two qualitatively different types of ion irradiations: (i) high-energy (100s keV -- MeV range) BIB inducing instant-separated cascades deeper in the bulk and (ii) low-energy (10s keV) FIB generating dense cascades in the vicinity of the surface, also varying ion masses. Both experiments allowed access to a broad spectrum of spatiotemporal resolution of radiation effects in \ce{Ga2O3}. 





\section*{\underline{\makebox[\textwidth][l]{Results}}}

\subsection*{Ion-irradiation induced competitive phase transition}

\begin{figure}[!hbt]
    \centering
    \includegraphics[width=\linewidth]{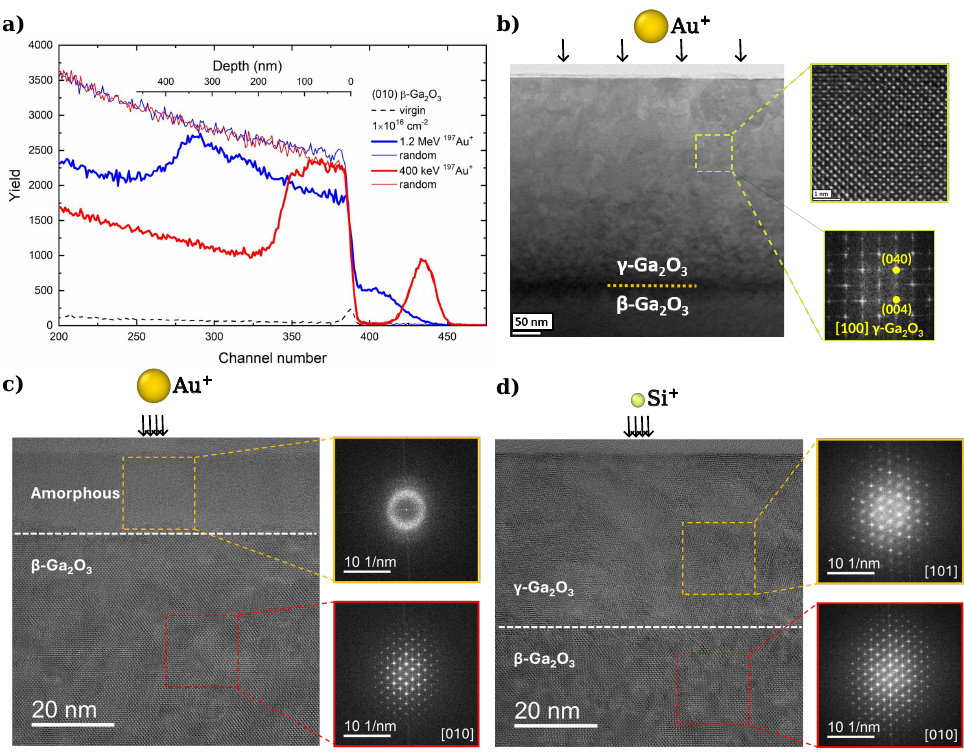}
    \caption{\textbf{Microstructural analysis of ion-implanted \bgao~ 
    samples via broad ion beam and focused ion beam irradiation} 
    \textbf{(a)} \ac{rbsc} spectra of (010) \bgao~ samples implanted with \ce{Au+} ions and a fluence of \qty{1e16}{\ions\per\centi\meter\squared}. Random (thin lines), unirradiated (dashed line), and channeling (thick lines) conditions at \qty{1.2}{\mega\electronvolt} and \qty{400}{\kilo \electronvolt}.
    \textbf{(b)} Low magnification \ac{bfstem} images of the sample implanted with \qty{1.2}{\mega\electronvolt} \ce{Au+} ions and \qty{1e16}{\ions \per\square\centi\meter}.
    The interface between the newly formed $\gamma$--phase and the pre-existing $\beta$--phase is indicated with dashed lines. The inset shows the magnified \ac{haadf}-STEM image showing the crystal structure of the sample post-implantation. A snippet shows the corresponding \ac{fft} of the inset.
        \textbf{(c)} \ac{hrtem} images after \qty{30}{\kilo\electronvolt} \ce{Au+} irradiation with a fluence of \qty{4e14}{\ions\per\centi\meter\squared}. 
    The \ac{fft} pattern of the amorphous layer from the corresponding region and the \ac{fft} pattern of the $\beta$ layer from the corresponding region along \hkl[010] zone axis were added.
    \textbf{(d)} \ac{hrtem} images after \qty{30}{\kilo\electronvolt} \ce{Si+} irradiation with a fluence of \qty{1e15}{\ions\per\centi\meter\squared}. 
    The \ac{fft} pattern of the $\gamma$ layer from the corresponding region along \hkl[101] zone axis and the \ac{fft} pattern of the $\beta$-layer from the corresponding region along \hkl[010] zone axis were added.
    }
    \label{fig:Figure1}
\end{figure}


Figure~\ref{fig:Figure1} shows the results of ion irradiation experiments using Au ions with energies 1.2\,MeV and 400\,keV (BIB) as well as for the Au and Si ions with energy 30\,keV (FIB). The \ac{rbsc} spectra in Figure~\ref{fig:Figure1}a and the transmission electron microscopy (TEM) image in Figure~\ref{fig:Figure1}b  demonstrate that under \qty{1.2}{\mega\electronvolt} BIB irradiation by Au ions the \bgao~structure irreversibly transforms into the defective spinel \ggao. The channeling spectrum obtained from the sample irradiated with these ions to a fluence of $1\times 10^{16}$ cm$^{-2}$ (thick blue line) shows a higher yield than that obtained from a pristine \bgao (dashed gray line), but still lower compared to the random one (thin blue line). This result unambiguously indicates that the lattice retains some crystallinity, attributed to the crystal structure of \ggao~without ever becoming amorphous. The corresponding TEM image in Figure~\ref{fig:Figure1}b also shows only two distinct regions with \ggao~ at the top and \bgao~ at the bottom. To further verify the crystallinity of the new phase, we show a higher magnification \ac{haadfstem} image of the transformed region, along with its Fast Fourier Transform (\ac{fft}) pattern. The \ac{fft} pattern confirms the crystallographic signature of the $\gamma$--phase.

To gain further insight into the phase transformation process in \ce{Ga2O3} under BIB irradiation conditions, we reduced the incident energy of Au ions to \qty{400}{\kilo\electronvolt}, while maintaining the same irradiation fluence. The ions are now deposited in a shallower end of the range, close to the surface. In Figure~\ref{fig:Figure1}a, the \ac{rbsc} spectrum for the sample irradiated with \qty{400}{\kilo \electronvolt} (thick red line) overlaps with the random spectrum measured for the same sample in a random direction (thin red line). This indicates that under this irradiation condition, the initial \bgao~lattice underwent a transition to the amorphous phase within approximately \qty{100}{\nano\meter} under the surface. Between 100 and \qty{150}{\nano\meter}, the \ac{rbsc} spectra is slightly lower than the random one, which we again attribute to a \ggao~  phase created underneath the amorphous layer. 

Furthermore, under the FIB irradiation condition, the sample irradiated with  \qty{30}{\kilo\electronvolt} Au ions shows only the amorphization of \bgao~ without any detectable transformation to \ggao,  as shown in Figure~\ref{fig:Figure1}c. In the \ac{haadfhrtem} images, two distinct regions can be clearly identified: a fully amorphous layer and the underlying crystalline \bgao. This interpretation is supported by the corresponding \ac{fft} patterns, which show the expected diffuse halo for the amorphous region and sharp reflections for the crystalline phase.
The fluence required to produce an amorphous layer of approximately \qty{20}{\nano\meter} thickness is \qty{4e14}{\ions\per\centi\meter\squared}, which is about 25 times lower than the fluence needed in the case of \qty{400}{\kilo\electronvolt} ion irradiation.
 \ac{fib} irradiation with Si ions of the same energy of \qty{30}{\kilo\electronvolt} did not lead to amorphization of the surface layer and, instead, a thick layer of \ggao~ formed, as can be seen in Figure~\ref{fig:Figure1}d. The corresponding \ac{fft} patterns for both \ggao~ and \bgao, shown in the same figure, display the characteristic diffraction maxima associated with each phase. Additional confirmation of the coexistence of amorphous and crystalline regions is provided by the \ac{ebsd} measurements included in the Supplemental Material (Figure~\ref{fig:EBSD_2}). Kikuchi patterns, which arise from the diffraction of electrons with the near-surface lattice structures reveal the lattice ordering of the most shallow layers. The comparison between patterns measured for un-irradiated and irradiated samples and the disappearance or transformation of some of the bands in those patterns is an indicator of a phase transformation occurring in the near surface layers. In Figure~\ref{fig:EBSD_2} we show the Kikuchi patterns for un-irradiated and irradiated samples of \bgao~with Si, Co and Au ions respectively. Si and Au irradiated patterns demonstrate the transmutation into \ggao~and amorphisation, respectively, in accordance to what is seen in figure \ref{fig:Figure1}. The pattern observed for the Co ions with 30 keV is less clear, while increasing the energy to 60 keV increases the sharpness of the lines indicating the formation of the \ggao.  

Unlike the BIB irradiation condition, which delivers ions on the surface of a millimeter-scale area homogeneously, the \ac{fib} irradiation reaches the same fluence in the same area as in the BIB experiments by gradually scanning the surface with only a sub-100 nm beam spot. 
This approach concentrates the impact point of the ions in a very small area, significantly increasing the ion flux at the same ion current. 

BIB irradiation was performed with a flux of approximately \qty{6.2E-03}{\ions\per\nano\meter\squared\per\second}, while the flux in our \ac{fib} experiments was \qty{7.9E+03} {\ions\per\nano\meter\squared\per\second}, which is six orders of magnitude higher. 
As a result, a much longer (about six orders of magnitude) time interval between successive displacement cascades is available for lattice recovery during the BIB irradiation than during the FIB experiments. 

To underpin the atom-level processes responsible for increased susceptibility to amorphization of the exceptional radiation hard material, we used a multiscale approach starting from a series of machine-learning-augmented molecular dynamics simulations of damage build up, which take place under conditions relevant to the current experiments, and then using a simulated annealing approach to extrapolating the physics of defect recombination far beyond the nanosecond time limits of molecular dynamics methods.

\subsection*{Atom-level insight on amorphization of $\beta$-\ce{Ga2O3} near the surface}


Traditionally, the ion mass is used as a parameter that can control the amorphization of irradiated materials. The heavier the ion, the denser the cascades it induces, subsequently leading to a higher concentration of defects produced and eventually to a loss of crystallinity. This effect was also observed in \ce{Ga2O3}\cite{Klevtsov2024},  therefore, we first examine whether irradiation of \bgao~ with heavier ions indeed enhances the susceptibility of this material to amorphization. The comparison of Figures~\ref{fig:Figure1}c and ~\ref{fig:Figure1}d, where we show the results of the \ac{fib} experiments with Au and Si ions, highlights the importance of the effect of the ion mass, which is associated with a higher number density of defects generated by heavy ions compared to lighter ones. However, the result shown in Figure~\ref{fig:Figure1}b does not follow the same trend. Here, Au ions with fluence even higher than in FIB experiments and much higher energy induced the crystal-to-crystal phase transition from \bgao~  to \ggao~without trace of the amorphous phase. It should be noted that an increase in ion energy generally leads to a reduction in defect production due to the decrease in non-ionizing energy losses in atomic collisions \cite{Wesch1993}. However, closer to the end of the ion range, the energy of the decelerated Au ions becomes comparable to that used in the \ac{fib} experiments. Hence, similar number densities of defects generated by every ion are expected, although at different depths, in both BIB and FIB experiments.

Furthermore, irradiation of \bgao with 400 keV Au ions also resulted in the formation of a thick amorphous surface layer, followed by a layer of \ggao. This finding again cannot be readily explained by models that attribute amorphization solely to the accumulation of radiation-induced defects. The 400 keV Au ion defect production profile also peaks beneath the surface, suggesting that amorphization, if related to the defect concentration, should first occur in a subsurface region. Instead, the experimentally observed amorphization originates at the surface. This behavior points to the existence of surface-mediated mechanisms that contribute significantly to the amorphization of gallium oxide under ion irradiation.


 \begin{figure*}[!hbt]
    \centering
    \includegraphics[width=\linewidth]{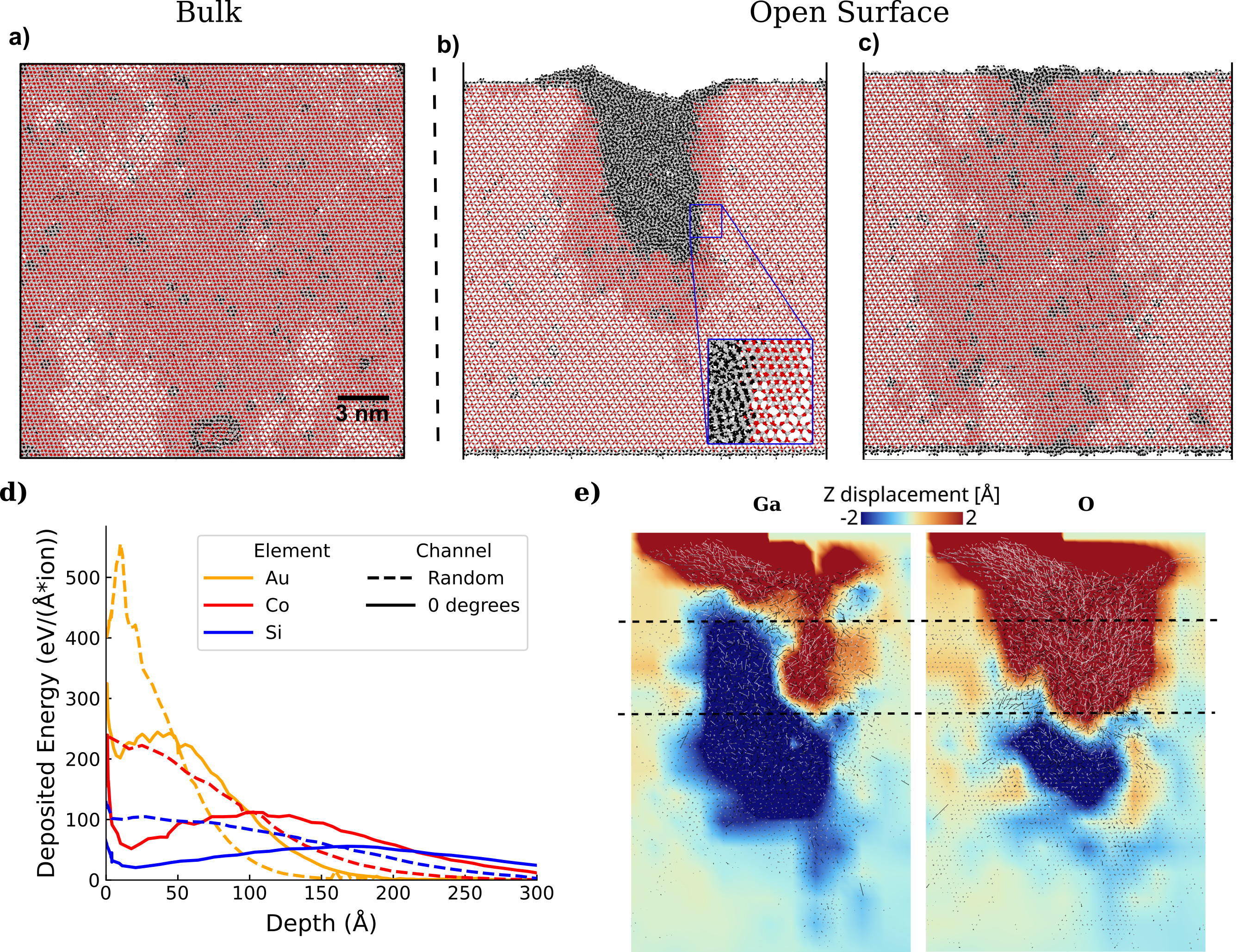}
    \caption{
    Simulation snapshots illustrating the results of the multiple cumulative impacts initiated in \bgao~by 30 keV Au ions. {\bf a)} shows the results of the bulk simulations for 200 impacts (see the main text for more detail), while {\bf b)} and {\bf c)} show the simulation results for \ac{fib} surface irradiation simulations for 100 impacts of Au and Co ions, respectively. 
    All snapshots show the 2 nm slices of the $\beta$-\ce{Ga2O3} taken near the center of the simulation cell in the $\beta \perp$ \hkl(100) direction (the reader's view). The scale bar shown in the snapshot in {\bf a)} is the same for all three snapshots.
    \textbf{d)} Depth profiles of \ce{Au}, \ce{Co} and \ce{Si} implanted with \qty{30}{\kilo\electronvolt} energy calculated using MDRANGE\cite{Nordlund1995}. 
    \textbf{e)} shows separately the analysis of preferential displacement ($z$-component) for Ga and O atoms in the simulation cell shown in {\bf b)}. Displacements are calculated with respect to the unirradiated positions and shown as the white (displacement toward the surface) and the black (displacement away from the surface) arrows. The background colour of the slice shows the average of the displacements (red and blue correspond to up and down displacements, respectively). The dashed lines are added for visual guidance of the region where the direction of the displacement of each species differs. } 
  \label{fig:simulations_BBvsFIB}
\end{figure*}

In order to gain insights on amorphization mechanisms of radiation-resistant \bgao~structure as well as to identify the factors playing leading roles in this process, we set up two types of ML-MD simulations in which the  Au ions with the same energy of \qty{30}{\kilo\electronvolt} initiated the cascades in \bgao~ at the open surface (periodic boundary conditions in lateral directions) imitating the \ac{fib} condition and in the bulk (periodic boundary conditions in all directions), which represents the 30 keV Au ions deep under the surface during the high energy BIB irradiation. The FIB condition was simulated by randomly selecting the entry point of an incoming ion within a squared area with a side of 5 nm to imitate the FIB beam spot, while in the bulk simulations the Au ions with 30 keV energy were introduced homogeneously everywhere in the simulation box using the box-shuffling technique, see the Method section.

In Figure~\ref{fig:simulations_BBvsFIB}a we show the simulation results of 300 cumulative impacts initiated by \qty{30}{\kilo\electronvolt} \ce{Au+} ions in the bulk, while Figure \ref{fig:simulations_BBvsFIB}b demonstrates the results after 100 cumulative impacts of the same ions but near the open surface. 

Each simulation snapshot shows a slice of the \bgao~  structure in the  direction $\beta \perp$ \hkl(100). (Here the latter refers to the direction perpendicular to the plane \hkl(100), that is not aligned with the \hkl[100] direction in the monoclinic \bgao). Since both $\beta$ and $\gamma$ phases of \ce{Ga2O3} share the same FCC lattice, while the Ga sublattices of both phases differ significantly, we emphasize the crystallinity of \ce{Ga2O3} by magnifying the FCC oxygen sublattice. In the figures, the oxygen atoms are shown in red, while the gallium atoms are gray. When the oxygen atoms lose their local FCC environment and become disordered (amorphous), their color changes to dark gray. All simulations were performed with an extra 60 ps of the relaxation run between the cascades, see the Methods section. This time is sufficient to reach a local minimum of a defective structure but not sufficient to recover the system to the most energetically stable configuration. 

We see that in the bulk simulations with three times as many ions as in the open surface simulations, the structure has mainly transformed into \ggao~with only occasional small disordered pockets. However, the same 30 keV Au ions hitting the open surface triggered amorphization of the irradiated area. Although the effective fluence in the open surface simulations is an order of magnitude higher than that in the bulk, we have previously shown \cite{azarov_universal_2023 , PhysRevLett.134.126101} that the structure transformed into a \ggao~phase is capable of withstanding very high fluences without losing its crystallinity. In Supplemental material, see Figure \ref{fig:accumulation}, we also show that \bgao~amorphizes under Au FIB irradiation already at a fluence 10 times lower.



The results presented in Figures \ref{fig:simulations_BBvsFIB}a and \ref{fig:simulations_BBvsFIB}b show a strong effect of the open surface on the susceptibility of \bgao~ to amorphization. As we see, the disordered oxygen sublattice did not recover the crystal structure near the open surface, even with additional simulation runs at elevated temperature to account for relaxation between the impacts. However, the FIB experiment with Si ions using the same irradiation energy and the same fluence did not result in amorphization of \bgao~ but in a phase transition to \ggao. To verify this result, we have also performed the simulations near the open surface with the ions much lighter than Au.

For computational efficiency, we used the Co ions instead of Si in our MD simulations. The Co ions generate damage closer to the surface than the Si ions, while they are still more than three times lighter than Au ions. To illustrate the efficiency of damage production by all three ions used in the present study, we plot in Figure~\ref{fig:simulations_BBvsFIB}d the deposited energy depth profiles for Au, Co and Si ions with the same energy of \qty{30}{\kilo\electronvolt}, obtained with the MDRANGE code \cite{Nordlund1995}. We plotted these distributions along and off the channeling direction aligned with the \hkl[-201] direction. One can see that the Au ions deposit energy much closer to the surface, in particular, when the direction of the ion impact is random. The impact in a random direction corresponds to later stages of irradiation when the channels are affected by the accumulated damage. Both Si and Co ions deposit energy in much flatter depth profiles, even in random directions.

Figure~\ref{fig:simulations_BBvsFIB}c shows that the Co ions did not initiate the amorphization of \bgao~ after the same number of consecutive impacts on the surface under FIB conditions, but triggered the transition to the $\gamma$-phase directly. With reference to Figure~\ref{fig:simulations_BBvsFIB}d, we anticipate that the Si ions are even less likely to amorphize \bgao~, which is confirmed in the analysis of the EBSD images shown in the Supplemental material; see Figure \ref{fig:EBSD_2} for all three ions, Au, Co and Si. In this figure, the EBSD images of the samples irradiated with Si and Co ions show the formation of the \ggao~ phase. Although the lines associated with \ggao~ are somewhat blurred, this can be explained by the small thickness of the newly formed phase layer. Increasing the Co ion energy to 60 keV, i.e. increasing the depth profile of the deposited energy, see Figure [Ask Kai to calculate it for you and add to Figure S1], resulted in a clear appearance of the $\gamma$ phase in a manner similar to that observed for Si ions. In contrast, there is no evidence of \ggao~ in the EBSD image of the sample irradiated with Au ions. Instead, the image exhibits a higher level of noise, suggesting amorphization of the surface layer. As we see in these simulations, the effect of the open surface on amorphization of \ce{Ga2O3}, demonstrated in Figure~\ref{fig:simulations_BBvsFIB}, is detrimental only in collaboration with the ion mass effect.

Since Ga and O atoms have different masses, preferential sputtering during the implantation of 30 keV ions may also explain the start of amorphous phase growth from the surface. Previously, we saw that high fluence Ga irradiation caused a stoichiometric imbalance at the end of the ion range, where the layer with the highest concentration of Ga atoms was amorphized \cite{azarov_universal_2023}. To verify this effect, we analyzed the stoichiometry of the composition throughout the simulation cells shown in Figures \ref{fig:simulations_BBvsFIB}a and \ref{fig:simulations_BBvsFIB}b and found that the stoichiometric imbalance caused by the different sputtering efficiency of two species is not sufficient to cause amorphization. As we show in the Supplemental material, Figure \ref{fig:stoichiometry_reversed}, the minimum value of $x$ in the formula \ce{Ga2Ox} does not fall below 2.8, while the cell has collapsed to an amorphous state at the values of $x < 2.5$. See the Supplemental material for more details. 

The effect of the open surface can be associated with the additional volume readily available when the generated defects relax in the direction of the surface, stabilizing the amorphous phase. Indeed, in the Supplemental Figure \ref{fig:amorph_E_V_curve} we show that the atomic volume in the amorphous phase is larger than the atomic volume for the crystalline \bgao~and \ggao~phases. Furthermore, repeated impacts by heavy ions at the surface transfer momentum to the atoms of the lattice in the direction of the ion beam. Hence, these atoms must mainly be displaced in the same direction. However, we notice that after multiple ion impacts, the atoms of both species involved in cascades become preferentially displaced in opposite directions, progressively driving the crystal away from its equilibrium lattice ordering. In Figure \ref{fig:simulations_BBvsFIB}e we show the same segment of the structure as in Figure \ref{fig:simulations_BBvsFIB}b but emphasize the $z$-component of the displacement vectors with respect to the unirradiated cell for Ga and O atoms separately. The white and black arrows correspond to upward and downward displacements, respectively, while the background color shows the local average of the vertical displacements (red is for "up" and blue for "down" displacements). The dashed lines are added to the figure to guide the eye while comparing the displacements of both Ga and O species. As one can see in this figure, the upward displacements of oxygen atoms in the outlined region overlap with downward displacements of gallium atoms. We observe that all atoms at shallower depths, regardless of their atomic species, were preferentially displaced toward the open surface, while the deeper located atoms were displaced towards the bulk. This is a consequence of heat spikes that are caused by heavy Au ions within a few nanometers of the surface. These heat spikes are characterized by an intensive chaotic motion of all atoms involved in the spike. The atoms within the heat spike have approximately similar kinetic energies that cause high pressure in the limited volume, which can be relieved toward the open surface. 

Although the motion is chaotic, we observe more pronounced displacements of the Ga atoms into the bulk following the initial direction of the momentum transfer, whereas the lighter O atoms preferentially move towards the surface. We explain this result by the difference in velocities of two atom species. The heavy Ga atoms move slower along the direction of the original momentum of the ion and get easily stuck in the cooler regions around the spike, whereas light oxygen atoms keep moving longer and eventually occupy the less dense regions close to the surface. 


To show that the observed behavior does not originate from the simulation artifact related to relaxation simulations at elevated temperature between the cascades, we show in the Supplemental material (Figure \ref{fig:cumulative_displacement}) that the preferential displacement of oxygen atoms is also observed in the simulation series of cumulative irradiation events where no additional annealing step was applied. Hence, this drift of oxygen atoms has a purely collision heat spike origin triggered by Au ions in the pre-surface regions. This conclusion is further supported by the analysis of displacements in the linear cascade regime during Co implantation (Figure \ref{fig:Co_displacements}). This analysis shows that in the linear-cascade mode, the displacements cannot be distinguished by atom species, all atoms follow the binary-type collisions resulting in random displacements of atoms of both types. 

\subsection*{Surface Effects on Damage Recovery under FIB Irradiation Condition}  \label{surface_effect}

\begin{figure*}[!hbt]
    \centering
    \includegraphics[width=\linewidth]{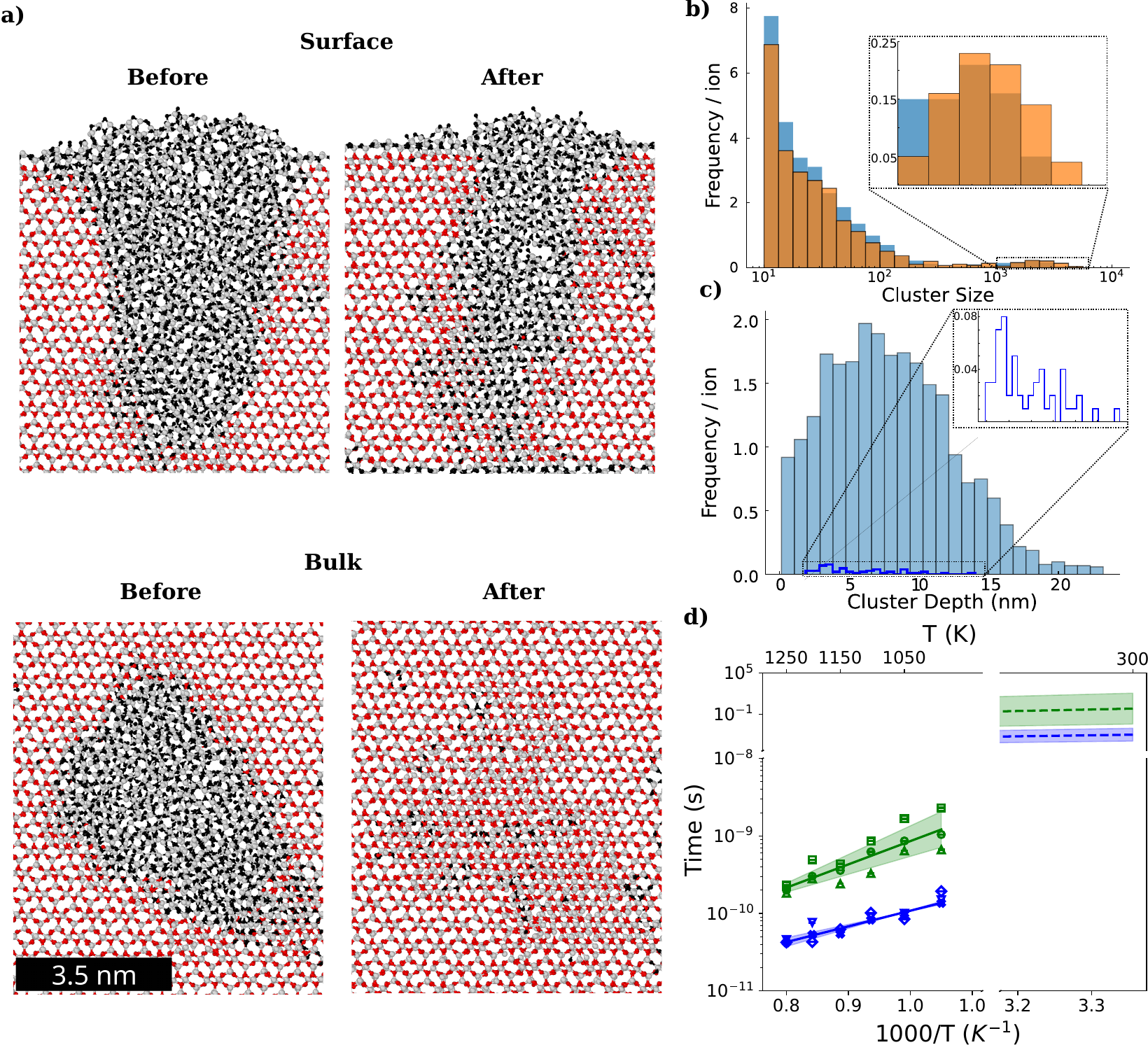}
    \caption{    
    \textbf{High-temperature relaxation results for \ce{Ga2O3} implanted with \ce{Au+} ions at \qty{30}{keV}}: 
    \textbf{(a)} Simulation snapshots showing the representative clusters of defects generated in the bulk ("Bulk") and near the surface ("Surface") by \ce{Au+} \qty{30}{\kilo\electronvolt} implantation before and after the annealing simulations for 100 ps at 1350 K in the NPT$\rightarrow$NVT$\rightarrow$NPT ensemble. The scale bar is the same for all four images. \textbf{(b)} Cluster size distributions formed by individual ion impacts within the bulk (blue) and at the surface (orange). The distributions are averaged over 100 simulations and clusters were determined by including non-FCC oxygen atoms within the cutoff of \qty{5}{\angstrom} as labeled by the \ac{ptm} analysis. The inset shows a zoom of the distribution for the largest clusters. \textbf{(c)} Depth distribution of defect clusters obtained in the open surface simulations. The solid line shows the depth distribution of the largest clusters highlighted in the inset of (b). \textbf{(d)} Recrystallization time versus temperature for open-surface (green lines) and bulk (blue lines) defects, corresponding to the time required for each to reduce to half of their initial dimensions. Each cluster’s temperature dependence was modelled with a separate Arrhenius-type function, allowing extrapolation of the recrystallization behaviour to room temperature 
    (indicated by dashed lines in the upper right corner).
    }
    \label{fig:Figure3}
\end{figure*}

In Figure~\ref{fig:simulations_BBvsFIB}c, we have already shown that ions with different mass produce distinct damage profiles. In principle, this difference is often used to explain the observed amorphization of gallium oxide by the heavy-mass ions. However, we also presented evidence that mass alone cannot explain the results: under different conditions, the same heavy ions with similar energies can induce either amorphisation or crystal-to-crystal transition from \bgao~to \ggao.
As we saw in the analysis above, the large defect clusters are produced under BIB and FIB implantation condition, the relaxation mechanisms and their characteristic timescales for these clusters may also depend on the distance from the surface. Hence, we performed additional analysis of the 100 individual cascade simulations and compared the results for bulk and surface configurations. For comparability, we used the Au ions of the same energy of 30 keV, which were introduced either just above the surface in the open-surface simulations or at random positions immediately outside the 9 nm diameter sphere in the middle of the bulk. In the former case, the ions were directed along the $z$-axis toward the simulation cell, whereas in the latter case they were directed toward the center of the sphere.


From the resulting structures obtained in these simulations, we selected cells containing large clusters of defects and extracted smaller subcells centered on these clusters to isolate them from the rest of the simulation cell and focus on recovery of a single cluster. The boundary conditions of the new cells were kept identical to those of the original cells. 
The new cells were allowed to evolve for 100 ps at 1350 K. An elevated temperature was used to accelerate the recovery dynamics, allowing the recovery process to be observed within the limited timescale of the MD simulations. 

In Figure~\ref{fig:Figure3}a we show the results of these simulations. Here, “Before” and “After” refer to the states of the simulation cells before and after annealing, respectively, whereas “Bulk” and “Surface” denote systems with fully periodic boundary conditions and with an open surface along the $z$-direction, respectively. The figure shows that, after identical relaxation runs, the defect cluster in the bulk fully recrystallizes from the amorphous state to the \ggao~phase. Although some residual defects remain, they are isolated point defects and do not form a cluster. In contrast, near the surface, only partial recrystallization is observed, and the majority of the damage remains intact. A similar surface effect has been reported in pure Sb systems, where the inability of the near-surface region to recrystallize was attributed to differences in atomic coordination relative to the crystalline phase \cite{Shen2023}. 

Furthermore, Figure~\ref{fig:Figure3}b compares the size distributions of the defect clusters generated by individual ions in the bulk and surface simulation cells. Only oxygen-related defects, i.e., non-FCC atoms located within a cutoff distance of \qty{5}{\angstrom} from one another, were included in the cluster analysis. The comparison shows that Au ions of the same energy generate smaller clusters of defects in the bulk (blue bars in Figure~\ref{fig:Figure3}b) than near an open surface (orange bars). In the latter case, large defect clusters form more frequently, as highlighted in the inset of Figure~\ref{fig:Figure3}b. The large clusters dissolve slower and have a chance to interact with a subsequent ion impact in their vicinity. Furthermore, the preferential formation of the largest clusters near the surface is evident in Figure~\ref{fig:Figure3}c, which compares the depth distribution of all clusters of defects in open-surface simulations with the depth distribution of the largest clusters ($> 1000$ defects), which is zoomed-in in the inset.



Since our goal is to compare the BIB and FIB conditions, it is important to note that the ion fluxes in these two regimes differ by approximately three orders of magnitude. Therefore, the effect of ion flux on the susceptibility of gallium oxide to amorphization must be taken into account.
To analyze whether there is sufficient time between the subsequent impacts for the disordered structure to recover the crystalline order, we performed a statistical study of the recrystallization dynamics of the damage. 
Using the method described in Figure~\ref{fig:Figure3}a, we isolated three representative large clusters into smaller subcells for the bulk and surface cases separately. These subcells with amorphous pockets were annealed at different temperatures, ranging from 1000 K to 1250 K. During these simulations, we monitored the number of "other" atoms excluding those at the surface using  \ac{ptm} (Polyhedral Template Matching) analysis with the value of the RMSD (Root-Mean-Square Deviation) cut-off of 0.15. 
To avoid overestimation of the defect concentration at high temperatures, we considered an atom to be a defect only if it had already been identified as such at the beginning of the annealing simulation. The simulations were terminated when the number of defects had decreased to half of its initial value.

In the experiment, the defect relaxation is expected to take place at 300 K. In Figure~\ref{fig:Figure3}d we extrapolate the Arrhenius plot obtained at elevated temperatures to 300 K taking into account the uncertainty indicated by the shadowed area (the change in the slope is due to the change in the logarithmic scale of the time axis). 
The results obtained show that the activation energy for recrystallization is much higher at the surface than that inside the bulk. This means that the extrapolated time for recrystallization to the room temperature will be much longer. More quantitatively, the large damage pockets vanish within milliseconds in bulk, while it takes many seconds for recrystallization of these pockets near the surface. The longer recovery times near the surface are attributed to the larger atomic volumes available near the surface for the atoms to stabilize in an amorphous phase (see Figure \ref{fig:amorph_E_V_curve}) and to the separation of atomic fluxes caused by thermal spikes near the surface (Figure \ref{fig:simulations_BBvsFIB}e). Separation owing to the different mobilities of both species progressively generates a stoichiometric imbalance with increasing fluence and thereby stabilizes the amorphous phase.

A simple estimation of the idle time between two consecutive spatially overlapping cascades can be obtained as follows. Assuming that all Au ions come to rest at the same depth and that the cascades induced by the \ce{Au+} ions overlap when the ions are within 20 nm, the cascade overlap area is A= \qty{314}{\nano\meter\squared}. The inverse of the flux provides the time interval between two overlapping cascades: $\Delta t  = (A\Phi)^{-1} = \qty{50}{ms}$ for Au ion current \qty{1}{\micro\ampere\per\centi\meter\squared}.
This value should be interpreted as an upper bound, used to estimate the strongest cascade overlap. However, even under this condition, the time between two consecutive cascades is from two to three orders of magnitude 
shorter than the estimated time obtained from our defect annealing simulations in the bulk. This means that the damage generated by cascades within the bulk material is very likely to recrystallize before the next cascade can strike in the vicinity of it. At the same time, the damage generated by the ions in the immediate vicinity of the surface recovers very slowly, and at room temperature this process can take tens of seconds, see the green dashed line in Figure \ref{fig:Figure3}e, which indicates the formation of very stable defects resulting in the amorphization of the surface layers $\beta$-\ce{Ga2O3} under Au ion irradiation in the FIB regime.

To summarize the discussion, we have found that dense collision cascades triggered by heavy ions, such as Au, in the vicinity of an open surface are likely to induce amorphization of the gallium oxide crystal structure. Thermal spikes promote ion mixing and, in the presence of a free surface, facilitate species separation by enhancing opposing atomic fluxes: oxygen atoms migrate toward the surface, whereas gallium atoms are driven into the bulk. The same cascades initiated much deeper beneath the surface does not lead to amorphization, as we see in the irradiation experiments with 1.2 MeV Au ions. Here the disordered \bgao~phase gradually transformed into the \ggao~phase. After that, the crystallinity was preserved until very high fluences of $10^{16}$ cm$^{-2}$. However, the experimental evidence of surface layer amorphization obtained with 400 keV Au irradiation is still counter-intuitive, since the result is expected to be closer to that of 1.2 MeV BIB irradiation rather than that of 30 keV FIB regime. 

Supplemental Figure \ref{fig:ranges_mdrange}a compares the ion ranges and depth profiles of nuclear energy deposition for 1.2 MeV and 400 keV Au ions in \bgao, as obtained from MDRANGE\cite{Nordlund1995}. Here we see that both curves are similar with the maxima of the ion range profiles at 200 and 70 nm, respectively. Although the maxima of the nuclear energy deposition are somewhat closer to the surface, at 120 nm and 40 nm, respectively, it is clear that the ions with the high implantation energy produce most of their damage deep beneath the surface. However, the RBS/C spectra in Figure \ref{fig:Figure1}a (thick red line) indicate that, at the same high fluence of $10^{16}$ ions cm$^{-2}$, 400 keV Au irradiation led to amorphization of \ce{Ga2O3}~ and that the amorphization originated at the sample surface. We attribute this behavior to the same mechanism observed for Au ions with energies below 50 keV. In a separate series of simulations, we identified this energy as an approximate upper limit for the formation of a single, nearly spherical thermal spike. At 60 keV, the collision cascades become elongated, indicating the onset of the subcascade-splitting regime \cite{Bac16}. 
To prove this, 
we performed MDRANGE simulations of 400 keV and 1.2 MeV Au ion implantation into \ce{Ga2O3} along the \hkl[-201] direction. Since only the fairly slow heavy Au ions can cause the dense heat spikes, we analyzed which fraction of the Au ions slow down to below 50 keV within the near-surface top 10 nm layer (and hence cause a near-surface heat spike).  For the 400 keV ions, we found that $0.08\pm0.01$\% of the ions slowed down to below 50 keV, whereas for 1200 keV,
this fraction was about two orders of magnitude lower, $\lesssim 0.001$\% (this upper limit was obtained from that in 100000 simulated 1200 keV ions, one slowed down to below 50 keV in the top 10 nm). This decrease at the higher energy is to be expected, since the nuclear collision cross section decreases strongly with energy above the maximum \cite{zbl}. 
These results show that, although the process is much less frequent than that in 30 keV Au FIB irradiation experiemnts, 400 keV Au ions can still occasionally generate near-surface thermal spikes capable of initiating amorphization from the surface into the bulk.

In conclusion, through a combination of experiments and simulations, we demonstrate that \ce{Ga2O3} undergoes amorphization under ion irradiation only when the ion mass and energy produce dense heat spikes in the immediate vicinity of a free surface. The presence of an open surface enables an asymmetric displacement of Ga and O atoms, leading to local non-stoichiometry. When the affected cascade volume is sufficiently large, this compositional imbalance suppresses recrystallization, thereby promoting amorphization. These findings, together with the quantitative analytical framework developed in this work, provide a basis for tailoring irradiation conditions to either prevent or induce amorphization, depending on the requirements of the intended application.

\section*{\underline{\makebox[\textwidth][l]{Experimental and simulation details}}}

\subsection*{Broad ion beam implantation of  in the MeV range}


To reach the condition for radiation damage buildup via defect accumulation in sparse collision cascades, we employ broad-beam ion irradiation of a \hkl[010]-oriented $\beta$--\ce{Ga2O3} wafer. In these experiments, we use \ce{Au+} ions with the energy of 1.2 MeV at room temperature at the dose of \qty{1e16}{\ions\per\centi\meter\squared} that is sufficient to cause phase transformation from $\beta$ to $\gamma$-\ce{Ga2O3} \cite{azarov_universal_2023}. 

The implantation was carried out with a 7$^\circ$ tilt from the surface normal to minimize channeling effects.
The microstructure of the implanted sample was characterized using a combination of \ac{rbsc} and \ac{stem}.
\ac{rbsc} measurements were performed with \qty{1.6}{\mega\electronvolt} \ce{He+} ions incident along the \hkl[010]  direction in a 165$^\circ$ backscattering geometry.
Cross‑sectional \ac{stem} analysis was conducted using a Cs‑corrected Thermo Fisher Scientific Titan G2 60–\qty{300}{\kilo\volt} microscope operated at \qty{300}{\kilo\volt}
An additional irradiation experiment was performed using \ce{Au+} ions at \qty{400}{\kilo\electronvolt} to achieve the same fluence of \qty{1e16}{\ions\per\centi\meter\squared}. In this case, only \ac{rbsc} measurements were conducted after irradiation.


\subsection*{Focused ion beam implantation of  in the keV range}

The impact of focused ion beam irradiation was studied using a commercial \hkl[-201]-oriented $\beta$--\ce{Ga2O3} wafers from Novel Crystal Technology Inc. 
The \ac{fib} irradiations were performed at the Ion Beam Center of the Helmholtz-Zentrum Dresden\,--\,Rossendorf using a commercial mass-separated Orsay CANION Z31Mplus \ac{fib} column and a Orsay Physics NanoSpace systems called TIBUSSII operating both with \ac{lmais}. 
Au-Si and Co-Nd alloy sources were used, and ions were accelerated at \qty{30}{\kilo\volt}.  
The desired ionic species were selected using a Wien filter according to their mass-to-charge ratio.
Double-charged ions are accelerated to an energy of \qty{60}{\kilo\electronvolt}.
The beam current was monitored with a Faraday cup and set to approximately \qty{10}{\pico\ampere}, which corresponds to an average time interval of \qty{16}{\nano\second\per\text{ion}} for singly charged ions. 
A spot size of approximately \qty{100}{\nano\meter} in diameter was achieved for an \ce{Au+} beam. 
In comparison, the lighter mass of \ce{Si+} and \ce{Co+} ions reduces the chromatic aberration, enabling these beams to be focused to a smaller spot size. 

A standard raster scanning pattern was employed for uniform irradiation. 
In this pattern, the beam scans each row pixel-by-pixel from the first to the last, then rapidly repositions to the beginning of the row directly below. 
Although the beam remains on during the flyback and repositioning between the pixels, the velocity is sufficiently high that the dose delivered during beam repositioning is negligible.
The square areas of \qty{5}{\micro\meter}~×~\qty{5}{\micro\meter} were irradiated with Au and Si ions, while the larger areas \qty{10}{\micro\meter}~×~\qty{10}{\micro\meter} were irradiated with Co, with all irradiations conducted at varying fluences, ranging from \qty{1e12}{\ions\per\centi\meter\squared} to \qty{8e15}{\ions\per\centi\meter\squared}.
The dwell time per pixel was adjusted so that each location was irradiated only once to reach the desired total fluence.


Cross-sectional \ac{tem} lamella preparation was carried out by in situ lift-out using a Thermo Fisher Helios 5 CX \ac{fib}-\acs*{sem} device. 
A carbon cap layer was deposited beginning with electron-beam-induced and followed by Ga-\ac{fib}-induced precursor decomposition to protect the sample surface. 
Afterwards, the \ac{tem} lamella was prepared using a \qty{30}{keV} Ga-\ac{fib} with adapted currents. 
It was transferred to a 3-post copper lift-out grid (Omniprobe) with an EasyLift EX nanomanipulator (Thermo Fisher). 
To minimize sidewall damage, \ce{Ga+} ions with \qty{5}{keV} energy were used for final thinning of the \ac{tem} lamella to electron transparency. 
\Ac{hrtem} images were acquired with an image-C$_s$-corrected Titan 80-300 microscope (FEI) operated at \qty{300}{kV}. 
\Ac{fft} analysis was done based on the recorded \ac{hrtem} micrographs. 

\subsection*{Machine-learning molecular dynamics simulations of ion beam irradiation }

The molecular dynamics method was used to study the effect of atomic species and the surface during ion irradiation of $\beta$--\ce{Ga2O3}. 
Interatomic interactions were modeled by the machine-learning interatomic potential developed for \ce{Ga2O3} systems in the tabGAP formalism \cite{Zhao2023}. 
This potential has been proven to offer computational results that are closely compared to the experimental ones \cite{azarov_universal_2023,Azarov2025,PhysRevMaterials.8.084601}.  
All simulations were performed using the Large-Scale Atomic/Molecular Massively Parallel Simulator (LAMMPS) \cite{LAMMPS}.

To ensure that all collision cascades triggered by the energetic ions develop within the central part of the simulation box, we selected a sufficiently large cell size 24.3$\times$24.7$\times$25 nm$^{3}$. 
First, an energy minimization of the input pure $\beta$--\ce{Ga2O3} was carried out using the conjugate gradient algorithm with an extra degree of freedom for cell size. 
This was followed by a thermal equilibration at \SI{300}{\kelvin} and \SI{0}{\bar} with an open surface using the isothermal-isobaric (NPT) ensemble on the periodic boundaries and the temperature and pressure damping parameters of 0.2 and \qty{2.0}{ps}, respectively. The system was equilibrated for \qty{20}{ps} of the simulated time. 
After that, the two bottom layers of the atoms were fixed to avoid lattice drift during simulations. In this configuration, the system was allowed to evolve for additional \qty{20}{ps} in the canonical (NVT) ensemble at \SI{300}{\kelvin}.

During ion impacts, the innermost atoms were switched from NVT to the microcanonical ensemble (NVE). 
Border cooling was applied to the layer of \SI{4}{\angstrom} near the periodic boundaries in the lateral directions and near the bottom of the cell in the layer of \SI{2}{\angstrom} above the fixed atoms. Figure \ref{fig:setup} shows a visual representation of the simulation setup. We note that during the irradiation simulations a small collective drift of the atoms in the cell appeared, which we eliminated by transforming the coordinates of the system back to the original center of mass before computing the displacement vectors.


The nuclear interaction of non-\ce{Ga} and \ce{O} ions with the lattice was described using the purely repulsive ZBL potential \cite{zbl}, while the high-energy interaction of \ce{Ga} and \ce{O} ions is already described by the \ce{Ga2O3} interatomic potential. 
In addition, to model electronic stopping, a friction force proportional to the velocity was applied to each moving atom in the lattice (that is, surrounded by 3 or more neighbours) above a given threshold of 10 eV \cite{SRIM}.
The ions were introduced at approximately 1 nm above the surface, beyond the cut-off point of the interatomic potential. 
To mimic the condition of the focused ion beam, we sampled the entry point of ions within a squared area of 5 $\times$ 5 nm$^{2}$ in the middle of the cell, as shown in Fig. \ref{fig:setup}. 
The direction of the ion was parallel to the Z direction, similarly to the experiment. Irradiation on the \hkl[-201] surface naturally prevents channelling.
Impact sequences were performed for the \ce{Au+} and \ce{Co+} ions. 
Simulations were carried out for \qty{15}{ps}, the time at which the damage is commonly assumed to be stable on MD timescales.

To emulate the high energy broad beam irradiation that has a damage peak well below the surface, we performed another set of simulations. Although 1.5 MeV Au ions produce more severe damage than 30 keV Au ions, for consistency between bulk and near-surface simulations, we initiate bulk cascades with the same energy of 30 keV, assuming that Au ions retain this energy deep in the bulk due to deceleration.
Hence, we setup simulations to analyse the damage produced by the same ions with the same energy, but without the presence of the open surface near the cascades. 
In this case, 300 \ce{Au+} ions with the energy of \qty{30}{keV} within the bulk $\beta$--\ce{Ga2O3} (i.e. using periodic boundary conditions in all three spatial dimensions) were consecutively inserted into the cell. Each ion was placed at a minimum distance of \qty{0.17}{nm} from all atoms in the cell within a spherical shell bounded by inner and outer radii of 9.0 and \qty{9.5}{nm}, respectively. 
The inserted ions were given an energy of 30 keV with the velocity vectors directed toward the centre of the simulation box to ensure a consistent impact geometry. This also guarantees a random orientation of the velocity vector with respect to the rest of the lattice. In a real implantation process, however, ions can strike any region of the bulk and approach from any direction. To reproduce this spatial randomness and prevent the accumulation of impacts in a single region, the simulation cell was randomly shifted between consecutive impacts. Consequently, the centre of the cell, and therefore the spherical region receiving the largest energy deposition, was different for each impact. 

Since in the experimental setups the cascades developed in the same location are separated by microsecond-to-second time intervals, we mimicked the possible lattice relaxation during these idle times at \SI{300}{\kelvin} by short simulated annealing runs at \SI{1300}{\kelvin} between impacts. 
During annealing, the entire system evolved in the NPT ensemble with the same relaxation constants as used in the relaxation simulation runs. 
The temperature was linearly increased  from 300 to \SI{1300}{\kelvin} during \qty{20}{ps}, followed by simulation at a constant temperature of \SI{1300}{\kelvin} for \qty{20}{ps}, after which the temperature was linearly decreased for another \qty{20}{ps} to \SI{300}{\kelvin}. 
When an open surface was present, the NPT ensemble was applied with rescaling of the degrees of freedom restricted to the $x$- and $y$-directions only. In the bulk case, the annealing step was applied only every 10 cascades, but for 200 ps, which makes it comparable to the open surface case. 

Molecular structures and defect analyses were performed using the OVITO visualisation tool \cite{ovito}. 
To detect defects,  PTM (Polyhedral Template Matching) analysis with the value of the RMSD (Root-Mean-Square Deviation) cut-off of 0.15 was performed on the oxygen sublattice. We identified a defect as an oxygen atom labelled as a non-\ac{fcc} structure. 

In the simulations, we sampled 100 overlapping ion impacts within a $5\times5\,\si{\square\nano\metre}$ region. 
This corresponds to the local ion density expected at a fluence of approximately \qty{4e14}{\ions\per\centi\meter\squared}. 
Although the full simulation cell is larger, the spatial extent of the collision cascades remains confined within this region because of the high density of atomic displacements. 
Therefore, we believe that this localized approach accurately captures the relevant damage processes while maintaining computational efficiency.


\section*{\underline{\makebox[\textwidth][l]{Acknowledgements}}}

Parts of this research was carried out at the Ion Beam Center at the Helmholtz-Zentrum Dresden – Rossendorf e. V., a member of the Helmholtz Association. 
The authors thank Andreas Worbs for his TEM lamella preparation expertise and his valuable and insightful discussions.
 T.F, F.D and Ru.H gratefully acknowledge the Research Council of Finland project SPATEC (grant nº 349690) for financial support and the CSC - IT Center for Science for computational resources as well as the Finnish Computing Competence Infrastructure (FCCI) for supporting this work with computational and data storage resources.
We acknowledge the M-ERA.NET Program for financial support via the GOFIB project (administrated by the Research Council of Norway project number 337627 in Norway, the Academy of Finland project number 352518 in Finland, and the tax funds on the basis of the budget passed by the Saxonian state parliament in Germany). 
U.B, N.K, G.H, T.F, A.K and F.D acknowledge support by the COST Action CA19140 FIT4NANO. J.G, A.A and A.K acknowledge acknowledge the DIOGO project (Research Council of Norway Project No. 351033) and the Norwegian Center for Transmission Electron Microscopy, NORTEM, supported by the Research Council of Norway project number 197405.

\section*{\underline{\makebox[\textwidth][l]{Disclosure Statement}}}

No potential conflict of interest was reported by the author(s).
\section*{\underline{\makebox[\textwidth][l]{Data Availability}}}

Data supporting the findings of this study are available from the corresponding author, T.F, upon reasonable request.

\newpage

\bibliographystyle{unsrt}
\bibliography{refs}

\appendix
\renewcommand{\thefigure}{S\arabic{figure}}
\setcounter{figure}{0}
\end{document}